\documentclass[prd,nofootinbib,showpacs,preprint]{revtex4}

\usepackage{graphicx} 
\usepackage{amsmath,amssymb} 
\usepackage{array} 

\usepackage[colorlinks]{hyperref}  
\usepackage[usenames,dvipsnames]{color} 

\newcommand{\vev}[1]{\langle #1 \rangle}

\begin{document}

\title{Gauge mediated supersymmetry breaking and the cosmology 
of Left-Right symmetric model}
\author{Sasmita Mishra} 
\email{sasmita@phy.iitb.ac.in}
\affiliation{Indian Institute of Technology, Bombay, Mumbai - 400076, India\\
and Indian Institute of Technology, Gandhinagar, Ahmedabad 382424}
\author{Anjishnu Sarkar\footnote{Address after October 1, 2008, Institute of Physics,
Bhubaneshwar 751005}} 
\email{anjishnu@phy.iitb.ac.in}
\affiliation{Indian Institute of Technology, Bombay, Mumbai - 400076, India}
\author{Urjit A. Yajnik}
\email{yajnik@phy.iitb.ac.in}
\affiliation{Indian Institute of Technology, Bombay, Mumbai - 400076, India\\
and Indian Institute of Technology, Gandhinagar, Ahmedabad 382424}

\date{}

\begin{abstract}
Left-Right symmetry including supersymmetry presents an important class
of gauge models which may possess natural solutions to many issues of 
phenomenology. Cosmology of such models indicates a phase transition 
accompanied by domain walls. Such walls must be unstable in order to not 
conflict with standard cosmology, and can further be shown to assist 
with open issues of cosmology such as dilution of unwanted relic densities 
and leptogenesis. In this paper we construct a model of gauge mediated 
supersymmetry breaking in which parity breaking is also signalled along
with supersymmetry breaking and so as to be consistent with cosmological 
requirements. It is shown that addressing all the stated cosmological
issues requires an extent of fine tuning, while in the absence of fine 
tuning, leptogenesis accompanying successful completion of the phase 
transition is still viable.

\end{abstract}

\maketitle

\section{Introduction}
\label{sec:intro} 
Left-Right symmetric model 
\cite{Pati:1974yy, Mohapatra:1974gc, Senjanovic:1975rk, Mohapatra:1980qe, Deshpande:1990ip}
has received considerable attention as a simple extension of 
the Standard Model (SM).
It provides elegant explanation of several open questions of SM
in addition to providing a natural explanation for the smallness of
neutrino masses\cite{Fukuda:2001nk, Ahmad:2002jz, Ahmad:2002ka, Bahcall:2004mz} 
via see-saw mechanism \cite{Minkowski:1977sc, Gell-Mann:1980vs, Yanagida:1979as, Mohapatra:1979ia}. 
In \cite{Lindner:1996tf,Raidal:2000ru} the possibility  that it be consistent 
as a TeV scale extension of the SM was explored.
In the supersymmetric setting it was shown by \cite{Deshpande:1992eu} that
$M_R\sim$ TeV is consistent with $SO(10)$ at $M_X\sim 10^{16}$GeV.
In this paper we adopt a similar approach viz., the scale of Left-Right
symmetry can be at TeV to PeV scale, and that supersymmetry (SUSY)
is preserved down to the same scale, so that the TeV / PeV scale 
is protected from unreasonable corrections from higher energy 
scales\footnote{For reviews of supersymmetry see 
\cite{Martin:1997ns, Aitchison:2005cf, Chung:2003fi}
and references therein.}. 

The cosmological phase transition accompanying the parity breakdown
is effectively a first order phase transition wherein domains of 
two different kinds of vacua separated
by a network of domain walls (DW) occur\cite{Kibble:1980mv}.
The need for ensuring a homogeneous Universe in late cosmology
requires the phase transition to end appropriately, in
particular that this network of domain walls be unstable.
The limits on the epoch to which the DW network may survive are
placed by the requirements of successful Big Bang Nucleosynthesis 
(BBN) and cosmic microwave background radiation (CMBR) data. 
A possible signature for a phase transition ending with decay
of domain walls at such energy scales is relic gravitational 
waves detectable at upcoming experiments\cite{Grojean:2006bp}.  

The unstable domain walls can have specific physical consequences.
One of the possibilities is that they are very slow in disappearing.
This is generically true if the difference in values of the effective 
potential across the wall is small. The energy density
of the domain wall complex scales with the cosmological scale factor 
$a$ as  $\rho \propto 1/a$, resulting in $a(t) \propto t^2$ leading
to a mild inflationary behaviour. This accords with a proposal of
thermal inflation\cite{Lyth:1995ka} (in some contexts called weak inflation) 
which can help to dilute unwanted relics arising in string 
theory\cite{Matsuda:2000mb}\cite{Kawasaki:2004rx}.

Unstable domain walls also provide the
non-adiabatic conditions for leptogenesis.
Theories with majorana neutrino masses and especially with gauged
$B-L$ symmetry present the interesting possibility of explaining
the baryon asymmetry of the Universe via thermal processes in
the early Universe\cite{Fukugita:1986hr}. This particular approach 
however has been shown to  generically require  the scale of  majorana neutrino 
mass, equivalently, the scale of gauged $B-L$ symmetry breaking 
in relevant models, to 
be $10^{11}$-$10^{13}$ GeV \cite{Davidson:2002qv, Hambye:2001eu}, 
with a more optimistic constraint $M_{B-L} >10^9$GeV 
\cite{Buchmuller:2003gz, Buchmuller:2004tu}. 
On the other hand, it has been shown \cite{Fischler:1990gn, Sahu:2004sb}
that the only real requirement imposed by Leptogenesis is that the presence
of heavy neutrinos should not erase lepton asymmetry generated by a given 
mechanism, possibly non-thermal. This places the modest bound $M_1> 10^4$GeV,
on the mass of the lightest  of the  heavy majorana neutrinos. 

There are two possible scenarios which exploit this window of low mass scale
for leptogenesis. One is the ``soft leptogenesis", 
\cite{Grossman:2003jv, D'Ambrosio:2003wy, Boubekeur:2004ez, Chun:2005ms}
relying on the decay of scalar superpartners of neutrino and a high degree
of degeneracy \cite{Pilaftsis:1997jf} 
in the mass eigenvalues due to soft SUSY  breaking terms. 
Another possibility for leptogenesis is provided by the unstable domain 
walls. This mechanism is analogous to that 
explored for the electroweak baryogenesis \cite{Cohen:1993nk},
provided a source for $CP$ asymmetry can be found. It has been shown 
\cite{Cline:2002ia} that the domain
walls occurring in models such as to be studied in this paper generically 
give spatially varying complex masses to neutrinos. Unstable domain walls 
of our models are therefore sufficient to ensure  the  required leptogenesis.

In an earlier work two of the authors have explored the possibility
of consistent cosmology in this class of models, primarily
the questions of removal of unwanted relics\cite{Sarkar:2007ic} and 
the role of domain
walls as possible catalysers of leptogenesis\cite{Sarkar:2007er}.
Exact Left-Right symmetry of the underlying model however does not permit 
instability of the domain walls.
In \cite{Sarkar:2007ic} it was shown that this circumstance is avoided
provided supersymmetry breaking in the hidden sector also breaks parity
symmetry and this is signalled through the soft terms.

Several proposals have been made for successful
phenomenological implementation of SUSY in a
Left-Right symmetric gauge theory \cite{Kuchimanchi:1993jg,Kuchimanchi:1995vk,
Aulakh:1997fq,Aulakh:1998nn,Aulakh:1997ba}. Implications of
gauge mediation of SUSY in the same was studied in
\cite{Mohapatra:1997kv,Chacko:1997en}.
More recently, this model has been studied in the context of the
LHC in \cite{Lee:2007aa,Babu:2008ep} and also a similar low energy
model in the context of cosmology in \cite{Shafi:2007xp}.
In all such models considered, spontaneous  gauge symmetry breaking 
required to recover SM phenomenology also  leads to observed parity 
breaking. However, for cosmological reasons  it is not sufficient 
to ensure local breakdown of parity. It has been  earlier proposed 
\cite{Yajnik:2006kc} that the occurrence of the SM like 
sector globally, i.e., homogeneously over the entire Universe, could be 
connected to the SUSY breaking effects  from the  hidden sector. 

Here we pursue this question in the context of two classes of models, 
one represented by Aulakh, Benakli, Melfo, Rasin and Senjanovic 
\cite{Aulakh:1997ba}\cite{Aulakh:1997fq} 
(ABMRS) and the other, proposed recently in Babu and Mohapatra, 
\cite{Babu:2008ep} (BM) both of which circumvent thorny issue of 
phenomenologically acceptable vacua by making minimal extensions of 
the basic scheme. The former uses two triplets $\Omega$ and $\Omega_c$
neutral under $U(1)_{B-L}$ while the latter uses one superfield $S$ 
singlet  under all the proposed gauge interactions as well as parity. 
In the approach we adopt, namely preservation of the Left-Right
symmetry to low energies, both these
models have a generic issue regarding occurance of domain walls
in cosmology which needs to be addressed. However the walls can
be easily removed without conflicting with phenomenology even with
very small parity breaking effects. In gauge mediated supersymmetry 
breaking GMSB, the SUSY breaking  effects are naturally small 
due to being  communicated at one loop and two loop orders. 
We propose the required (GMSB) scheme  to obtain natural conditions 
for SUSY breaking and also for the disappearance of the domain walls.  
We show that the smallness migrates into parity breaking and is 
adequate to ensure global choice of true ground state in the Universe.

In sec. \ref{sec:mslrm} we discuss the two models ABMRS and BM.
In sec. \ref{sec:csmlgyBrk} we discuss the soft terms arising
in these models and the constraints imposed by consistent cosmology;
presenting some of the concerned formulae in the appendix \ref{sec:ap1}.
In sec \ref{sec:gmsb} we pursue the consequence of implementing 
generic GMSB in these theories. In sec. \ref{sec:customgmsb} we 
propose a modification 
of the scheme to implement combined breaking of SUSY as well 
as parity in a manner consistent with low energy symmetries 
and successful cosmology. Section \ref{sec:cnclsn} contains
discussion of our results.

\section{Minimal Supersymmetric Left-Right Model : A recap}
\label{sec:mslrm} 

The quark, lepton and Higgs fields for the minimal left-right SUSY
model, with their respective quantum numbers under the gauge group
$SU(3)_c$ $\otimes ~SU(2)_L$ $\otimes ~SU(2)_R$ $\otimes ~U(1)_{B-L}$ are
given by,
\begin{eqnarray}
Q = (3,2,1,1/3), & \quad & Q_c = (3^*,1,2,-1/3), \nonumber  \\
L = (1,2,1,-1),  & \quad & L_c = (1,1,2,1), \nonumber \\
\Phi_i = (1,2,2,0),    & \quad & \textrm{for } i = 1,2, \nonumber \\
\Delta = (1,3,1,2),    & \quad & \bar{\Delta} = (1,3,1,-2), \nonumber \\
\Delta_c = (1,1,3,-2), & \quad & \bar{\Delta}_c = (1,1,3,2). 
\label{eq:minimalset}
\end{eqnarray}
where we have suppressed the generation index for simplicity of notation.
In the Higgs sector, the bidoublet $\Phi$ is doubled to have non-vanishing 
CKM matrix, whereas the $\Delta$ triplets are doubled to have anomaly 
cancellation.  
Under discrete parity symmetry the fields are prescribed to transform as,
\begin{eqnarray}
Q \leftrightarrow Q_c^*, \quad & 
L \leftrightarrow L_c^*, \quad & 
\Phi_i \leftrightarrow \Phi_i^\dagger,  \nonumber \\
\Delta \leftrightarrow \Delta_c^*,  \quad & 
\bar{\Delta} \leftrightarrow \bar{\Delta}_c^*.
\label{eq:parity} 
\end{eqnarray}
However, this minimal left-right symmetric model is unable
to break parity spontaneously \cite{Kuchimanchi:1993jg, Kuchimanchi:1995vk}.
Inclusion of nonrenormalizable terms gives a more realistic structure of
possible vacua \cite{Mohapatra:1995xd}\cite{Aulakh:1998nn,Aulakh:1997fq}. 
Such terms were studied for the case when the scale 
of $SU(2)_R$ breaking is high, close to Planck scale. We shall not 
pursue this possibility here further, retaining interest in TeV to PeV
scale phenomenology.

\subsection{The ABMRS model with a pair of triplets}
Due to difficulties with the model discussed above, an early model to 
be called minimal by its authors is ABMRS\cite{Aulakh:1997ba,
Aulakh:1998nn,Aulakh:1997fq}. Here two triplet 
fields $\Omega$ and $\Omega_c$, were added, with the following quantum numbers,
\begin{equation}
\Omega = (1,3,1,0), \quad \Omega_c = (1,1,3,0),
\end{equation} 
which was shown to improve the situation with only the renormalisable
terms\cite{Aulakh:1997ba, Aulakh:1997fq, Aulakh:2003kg}.
It was shown that this model breaks down to minimal 
supersymmetric standard model (MSSM) at low scale.
This model was studied in the context of cosmology in 
\cite{Yajnik:2006kc, Sarkar:2007ic} specifically, the mechanism for 
leptogenesis via Domain Walls in \cite{Sarkar:2007er}.

The superpotential for this model is given by,
%
\begin{eqnarray}
 W_{LR}&=& {\bf h}_l^{(i)} L^T \tau_2 \Phi_i \tau_2 L_c 
 + {\bf h}_q^{(i)} Q^T \tau_2 \Phi_i \tau_2 Q_c 
 +i {\bf f} L^T \tau_2 \Delta L 
 +i {\bf f} L^{cT}\tau_2 \Delta_c L_c \nonumber \\
&& + ~m_\Delta  {\rm  Tr}\, \Delta \bar{\Delta} 
  + m_\Delta  {\rm Tr}\,\Delta_c \bar{\Delta}_c
  + \frac{m_\Omega}{2} {\rm Tr}\,\Omega^2 
  + \frac{m_\Omega}{2} {\rm Tr}\,\Omega_c^2 \nonumber \\
&& + ~\mu_{ij} {\rm Tr}\,  \tau_2 \Phi^T_i \tau_2 \Phi_j 
  + a {\rm Tr}\,\Delta \Omega \bar{\Delta}
  + a {\rm Tr}\,\Delta_c \Omega_c \bar{\Delta}_c \nonumber \\
& &  + ~\alpha_{ij} {\rm Tr}\, \Omega  \Phi_i \tau_2 \Phi_j^T \tau_2 
  + \alpha_{ij} {\rm Tr}\, \Omega_c  \Phi^T_i \tau_2 \Phi_j \tau_2 ~.
\label{eq:superpot}
\end{eqnarray} 
Since supersymmetry is broken
at a very low scale, we can employ the $F$ and $D$ flatness conditions
obtained from the superpotential, to get a possible solution for the 
vacuum expectation values (vev's)
for the Higgs fields. 
\begin{equation} 
\begin{array}{ccc}
\langle \Omega \rangle = 0, & \qquad
\langle \Delta \rangle = 0, & \qquad
\langle \bar{\Delta} \rangle = 0, \\ [0.25cm]
\langle \Omega_c \rangle = 
\begin{pmatrix}
\omega_c & 0 \\
0 & - \omega_c
\end{pmatrix}, & \qquad
\langle \Delta_c \rangle =
\begin{pmatrix}
0 & 0 \\
d_c & 0
\end{pmatrix}, & \qquad
\langle \bar{\Delta}_c \rangle =
\begin{pmatrix}
0 & \bar{d}_c \\
0 & 0
\end{pmatrix}.
\label{eq:rhvev} 
\end{array}
\end{equation}
This solution set is of course not unique. Since the original theory is
parity invariant a second  solution for the $F$ and $D$ flat
conditions exists, with Left type fields' vev's exchanged with those 
of the Right type fields \cite{Sarkar:2007ic, Sarkar:2007er}. 

With vev's as in eq. (\ref{eq:rhvev}) the pattern of breaking is 
\begin{eqnarray}
SU(2)_L \otimes SU(2)_R \otimes U(1)_{B-L} &\stackrel{M_R}{\longrightarrow}& 
SU(2)_L \otimes U(1)_R \otimes U(1)_{B-L}\\
&\stackrel{M_{B-L}}{\longrightarrow}& SU(2)_L \otimes U(1)_Y
\end{eqnarray}
It was observed \cite{Aulakh:1997fq} that supersymmetric breaking
imposes a condition on the scales of breaking, with respect to the electroweak 
scale $M_W$,
\begin{equation}
M_R M_W \simeq M_{B-L}^2
\end{equation}
This relation raises the interesting possibility that the scale
of $M_R$ can be as low as $10^4$ to $10^6$ GeV, with corresponding
very low scale $10^3$ to $10^4$GeV of lepton number violation, 
opening the possibility of low energy leptogenesis
\cite{Sahu:2004sb}\cite{Sarkar:2007er}.

\subsection{The BM model with a single singlet}
An independent approach to improve the minimal model with
introduction of a parity odd singlet \cite{Cvetic:1985zp}, was adopted
in \cite{Kuchimanchi:1993jg, Kuchimanchi:1995vk}.
However this was shown at tree level to lead to charge-breaking vacua 
being at a lower potential than charge-preserving vacua.

Recently, an alternative to this has been considered in \cite{Babu:2008ep} 
where a  superfield $S = (1,1,1,0)$ also singlet under parity is included
in addition to the minimal set of Higgs required as in eq. (\ref{eq:minimalset}).
The superpotential is  given by,
\begin{center} 
$ W_{LR} = W^{(1)} + W^{(2)}$
 \end{center} 
Where
\begin{eqnarray} 
 W^{(1)} &=& {\bf h}_l^{(i)} L^T \tau_2 \Phi_i \tau_2 L_c 
 + {\bf h}_q^{(i)} Q^T \tau_2 \Phi_i \tau_2 Q_c\nonumber\\ 
&& + ~ i {\bf f^\ast} L^T \tau_2 \Delta L 
 + ~i {\bf f} L^{cT}\tau_2 \Delta_c L_c \nonumber \\
&& + ~ S~[~\lambda^\ast ~ {\rm Tr}\,\Delta  \bar{\Delta}
  + ~ \lambda ~ {\rm Tr}\,\Delta_c  \bar{\Delta}_c
  + ~ \lambda^\prime_{ab} {\rm Tr}\, \Phi_a^T \tau_2 \Phi_b \tau_2 - M_R^2~]
\label{eq:Wone}\\
W^{(2)} &=& ~M_\Delta  {\rm  Tr}\, \Delta \bar{\Delta} 
  + M_\Delta^\ast  {\rm Tr}\,\Delta_c \bar{\Delta}_c\nonumber\\
&& + ~\mu_{ab} {\rm Tr}\, \Phi^T_a \tau_2 \Phi_b \tau_2
   + M_s S^2 + \lambda_s S^3
\label{eq:Wtwo}
 \end{eqnarray} 
For a variety of phenomenological reasons\cite{Babu:2008ep}, the terms 
in $W^{(2)}$ may be assumed to be zero. The presence of linear terms in 
$S$ in $W^{(1)}$ makes possible the following SUSY vacuum,
\begin{equation}
\vev{S} = 0, \quad \lambda v_R \bar{v}_R + \lambda^\ast
v_L \bar{v}_L = M_R^2
\label{eq:SSvev} 
\end{equation}
In the ABMRS model, the introduction of a separate $\Omega$ field for 
each of the sectors $L$ and $R$ permits local preference of one 
sector over the other through spontaneous symmetry breaking. The preference
however remains strictly local. Another region of space  can equally
well have vacuum of another type. In the BM model however, due to the
$S$ field being neutral including under parity, such a distinction 
cannot arise even locally. This is reflected in the above equation 
(\ref{eq:SSvev}) where we have a flat direction in 
the $v_L$ - $v_R$ space 
\footnote{
The first equation of sec. 3, of 
\cite{Babu:2008ep} is a special case of our condition 
(\ref{eq:SSvev})
}.
A more general treatment of the possible vacua is included in appendix 
\ref{sec:ap1}.
Nevertheless, 
\begin{equation}
v_L=\bar{v}_L=0, \qquad |v_R|=|\bar{v}_R|=\frac{M_R}{\sqrt{\lambda}}
\label{eq:SSLsol}
\end{equation}
is a possible vacuum\cite{Babu:2008ep}
in which we recover the known phenomenology.

The important result is that after SUSY breaking and emergence of 
SUSY breaking soft terms,  integrating out heavy sleptons modifies
the vacuum structure due to Coleman-Weinberg type one-loop terms 
which must be treated  to be of same order as the other terms in 
$V^{eff}$. Accordingly, it is shown\cite{Babu:2008ep} that the 
$V^{eff}$ contains terms of the form
\begin{equation} 
 V_{1-loop}^{eff} (\Delta_c) \sim - |f|^2 m_{L^c}^2
{\rm  Tr}\, (\Delta_c \Delta_c^\dagger) A_1^R - |f|^2 m_{L^c}^2
{\rm  Tr}\, (\Delta_c \Delta_c)
{\rm  Tr}\, (\Delta_c^\dagger \Delta_c^\dagger) A_2^R
\label{eq:onelR}
 \end{equation} 
where $A_1^R$ and $A_2^R$ are constants obtained from expansion of the
effective potential.
Presence of these terms is shown to lead to the happy consequence 
of a preference  for electric charge preserving vacuum over the 
charge breaking  vacuum, provided $m_{L^c}^2 < 0$.

For the purpose of the present paper it is important to note that
even assuming that some soft terms will lift the flat direction 
in (\ref{eq:SSvev}), we still have no source of breaking $L$-$R$ symmetry.
This means that
\begin{equation}
v_R=\bar{v}_R=0, \qquad |v_L|=|\bar{v}_L|=\frac{M_R}{\sqrt{\lambda^*}}
\label{eq:SSRsol}
\end{equation}
also constitutes a valid solution of eq. (\ref{eq:SSvev}). In this vacuum
the soft terms can give rise to the following terms in the effective 
potential,
\begin{equation} 
 V_{1-loop}^{eff} (\Delta) \sim - |f|^2 m_L^2
{\rm  Tr}\, (\Delta \Delta^\dagger) A_1^L - |f|^2 m_L^2
{\rm  Tr}\, (\Delta \Delta) {\rm  Tr}\, (\Delta^\dagger\Delta^\dagger)A_2^L
\label{eq:onelL} 
\end{equation} 
with $A_1^L$ and $A_2^L$ constants.
Thus the choice of known phenomenology is only one of two possible local
choices, and formation of domain walls is inevitable.

\section{Cosmology of breaking and Soft terms}
\label{sec:csmlgyBrk} 
Due to the existence of two different sets of solutions for the possible 
vev's, formation of domain walls (DW) is inevitable \cite{Sarkar:2007ic, Sarkar:2007er}
in both the models considered above. Stable walls are known to overclose
the Universe\cite{Kibble:1980mv}\cite{Hindmarsh:1994re} and are undesirable.
However, a small inequality in the free energies in the vacua on the
two sides of the walls is sufficient to destabilize them. The lower bound
on the temperature upto which such walls may persist in the Universe is
set by the known physics of BBN, $T\sim 1 MeV$. Let the temperature by which
the wall complex has substantially decayed, in particular has ceased to
dominate the energy density of the Universe, be $T_D$. 
It has been estimated that the free energy density difference $\delta \rho$ 
between the vacua which determines the pressure difference across a domain wall 
should be of the form \cite{Vilenkin:1984ib}\cite{Preskill:1991kd} 
\begin{equation}
\delta \rho \sim T_D^4
\label{eq:DWtemp} 
\end{equation} 
in order for the DW to disappear at the scale $T_D$.

It has been observed in \cite{Rai:1992xw} that parity breaking
effects suppressed by Planck scale are sufficient to remove
the DW. Black Holes can carry only locally conserved charges, 
and therefore there would exist processes in Quantum Gravity 
which do not conserve global charges as well as violate discrete 
symmetry such as parity. Thus we expect parity breaking terms 
induced by quantum gravity in the effective lagrangian.
Assuming that the pressure across the walls is created by terms 
such as $C\varphi^6/M_{Pl}^2$, it is easy to check that reasonable
range of values of the order parameter $\langle\varphi\rangle$ and 
$T_D$ exist for which the walls can disappear without conflicting with cosmology.
This is especially true of high scale models, $M_R\gtrsim 10^{11}$GeV.
We shall be interested in a low (PeV) scale model where Planck scale
effects can be ignorable and where the parity breaking should arise
from known effects which can be counterchecked against other observables.

In ABMRS model at the scale $M_R$, $SU(2)_R\otimes ~U(1)_{B-L}$ 
breaks to $U(1)_R$ $\otimes ~U(1)_{B-L}$, equally well, 
$SU(2)_L\otimes ~U(1)_{B-L}$ breaks 
to $U(1)_L$ $\otimes ~U(1)_{B-L}$ depending on
which of the two $\Omega$ fields acquires a vev. Thus domain walls are formed
at the scale $M_R$. At a lower scale $M_{B-L}$ when the Higgs triplet
$\Delta$'s get vev, $U(1)_R\otimes ~U(1)_{B-L}$ breaks to $U(1)_Y$
or $U(1)_L$ $\otimes ~U(1)_{B-L}$ breaks to $U(1)_{Y'}$.
In BM model the breaking is directly to $SU(2)_L\otimes U(1)_Y$
or equally well, $SU(2)_R\otimes ~U(1)_Y'$. As per the analysis reported
above we are assuming that the $S$ field does not acquire a vev.
Thus in each of the models we have MSSM after the $M_{B-L}$ or 
the $M_R$ scale, respectively.
SUSY breaking soft terms emerge below the SUSY breaking scale $M_S$.
We now proceed with the stipulation advanced in \cite{Yajnik:2006kc} that
the role of the hidden sector dynamics is not only to break SUSY but 
also break parity. This permits in principle a relation between observables
arising from the two apparently independent breaking effects.

The soft terms which arise in the two models ABMRS and BM may be
parameterized as follows 
\begin{eqnarray}
\mathcal{L}_{soft}^1 &=& ~m_1^2 \textrm{Tr} (\Delta \Delta^{\dagger}) +
m_2^2 \textrm{Tr} (\bar{\Delta} \bar{\Delta}^{\dagger})\nonumber\\
&& + ~m_3^2 \textrm{Tr} (\Delta_c \Delta^{\dagger}_c) +
m_4^2 \textrm{Tr} (\bar{\Delta}_c \bar{\Delta}^{\dagger}_c)
\label{eq:eqnsoftone}\\
\mathcal{L}_{soft}^2 &=&
\alpha_1 \textrm{Tr} (\Delta \Omega \Delta^{\dagger}) +
\alpha_2 \textrm{Tr} (\bar{\Delta} \Omega \bar{\Delta}^{\dagger})\nonumber\\
&& + ~\alpha_3 \textrm{Tr} (\Delta_c \Omega_c \Delta^{\dagger}_c) +
\alpha_4 \textrm{Tr} (\bar{\Delta}_c \Omega_c \bar{\Delta}^{\dagger}_c)
\label{eq:eqnsofttwo}\\
\mathcal{L}_{soft}^3 &=&
 ~\beta_1 \textrm{Tr} (\Omega \Omega^{\dagger}) +
\beta_2 \textrm{Tr} (\Omega_c \Omega^{\dagger}_c)
\label{eq:eqnsoftthree}\\
\mathcal{L}_{soft}^4 &=&
S[\gamma_1 \textrm{Tr} (\Delta\Delta^{\dagger}) +
\gamma_2 \textrm{Tr} (\bar{\Delta}\bar{\Delta}^{\dagger})]\nonumber\\
&& + ~ S^*[\gamma_3 \textrm{Tr} (\Delta_c\Delta^{\dagger}_c) +
\gamma_4 \textrm{Tr} (\bar{\Delta}_c\bar{\Delta}^{\dagger}_c)]
\label{eq:eqnsoftfour}\\
\mathcal{L}_{soft}^5 &=& ~  {\tilde\sigma}^2 |S|^2
\label{eq:eqnsoftfive} \end{eqnarray}
%
For ABMRS model the relevant soft terms are given by,
\begin{equation} 
\mathcal{L}_{soft} = \mathcal{L}_{soft}^1 + \mathcal{L}_{soft}^2
+ \mathcal{L}_{soft}^3
 \end{equation}  
For BM model the soft terms are given by,
\begin{equation} 
\mathcal{L}_{soft} = \mathcal{L}_{soft}^1  + \mathcal{L}_{soft}^4
+ \mathcal{L}_{soft}^5
 \end{equation}  
Using the requirement of eq. (\ref{eq:DWtemp}) we can constrain the 
differences between the soft terms in the Left and Right sectors 
\cite{Sarkar:2007ic, Sarkar:2007er}. According to eq. (\ref{eq:SSvev}) 
$S$ field does not acquire a vev in the physically relevant vacua
and hence the terms in eq.s (\ref{eq:eqnsoftfour}) and (\ref{eq:eqnsoftfive}) 
do not contribute to the vacuum energy. The terms in eq. (\ref{eq:eqnsofttwo})
are suppressed in magnitude relative to those in eq. (\ref{eq:eqnsoftthree})
due to having $\Omega$ vev's to one power lower. This argument assumes
that the magnitude of the coefficients $\alpha$ are such as to not mix
up the symmetry breaking scales of the $\Omega$'s and the $\Delta$'s.

To obtain orders of magnitude we have taken the $m_i^2$ parameters 
to be of the form $m_1^2 \sim m_2^2$
$\sim m^2$ and $m_3^2 \sim m_4^2$ $\sim m^{'2}$ \cite{Sarkar:2007er} 
with $T_D$ in the range $ 10 - 10^3$ GeV \cite{Kawasaki:2004rx}. 
For both the models we have taken the value of the $\Delta$ vev's as 
$d \sim 10^4$ GeV. For ABMRS model additionally we take $\omega \sim 10^6 $ GeV.
The resulting differences required for successful removal of domain walls
are shown in Table \ref{tab:DWalls}. 

\begin{table}
\begin{center} 
{\setlength\extrarowheight{1.5mm}
\begin{tabular}{cc|c|c|c} 
\hline
$T_D/$GeV & $\sim$ & $10$ & $10^2$ & $10^3$ \\ \hline \hline
$(m^2 - m^{2\prime})/\mbox{GeV}^2$ & $\sim$ &
$10^{-4}  $  & $1$  & $10^{4} $\\ [1mm]
$(\beta_1 - \beta_2)/\mbox{GeV}^2 $ & $\sim$ &
$10^{-8}  $ & $10^{-4} $ & 
$1$ \\ [1mm] \hline \hline
\end{tabular} }
\end{center}
\caption{Differences in values of soft supersymmetry breaking parameters 
for a range of domain wall decay temperature values $T_D$. The 
differences signify the extent of parity breaking. 
}
\label{tab:DWalls}
\end{table}
We see from table \ref{tab:DWalls} that assuming both the mass-squared 
differences $m^2-m'^2$ and $\beta_1-\beta_2$ arise from the same dynamics,
$\Omega$ fields are the determinant of the cosmology. This is because
the lower bound on the wall disappearance temperature $T_D$ required
by $\Omega$ fields is higher and the corresponding $T_D$ is reached 
sooner.  This situation changes if for some reason $\Omega$'s do not
contribute to the pressure difference across the walls. The BM
model does not have $\Omega$'s and falls in this category. 

During the period of time in between destabilization of the DW and their
decay, leptogenesis occurs due to these unstable DW as discussed in
\cite{Cline:2002ia, Sarkar:2007er}. After the disappearance of the walls at
the scale $T_D$, electroweak symmetry breaks at a scale $M_{EW} \sim 10^2$ 
GeV and standard cosmology takes over.
In the next section we discuss the implementation of GMSB scenario 
for these models.

\section{Gauge mediated SUSY breaking in minimal supersymmetric models} 
\label{sec:gmsb} 

In gauge mediated SUSY breaking proposal, dynamical breaking of
SUSY in a hidden sector is communicated to visible sector
through a field $X$ singlet under visible gauge interactions,
and one or more conjugate pairs of chiral superfields called
messenger fields which together constitute a vector like, 
anomaly free representation
\cite{Dine:1993yw, Dine:1994vc,Dine:1995ag}. For reviews the 
reader may refer \cite{Weinberg:2000cr,Dubovsky:1999xc, Martin:1997ns}.
The choice of charges for the messenger fields ensures that they 
do not spoil gauge coupling unification. A simple choice is to
choose them to be complete representations of the possible
grand unification group, which in the Left-Right symmetric case 
is $SO(10)$.

The dynamical SUSY breakdown causes the messenger fields to 
develop SUSY violating interactions with the
hidden sector, while they also interact with the (s)quarks, 
(s)leptons and Higgs(inos) 
via gauge and gauginos interactions. Gaugino fields get a mass 
at one loop due 
to these interactions, while gauge invariance prevents gauge fields 
from acquiring any mass. As such supersymmetry is broken in the 
visible sector. 

We denote the messenger sector fields to be $\Phi_i$ and its complex
conjugate representation to be $\bar{\Phi}_i$, with $i$ indexing 
the set of several possible messengers. These couple to a chiral
superfield singlet $X$ via yukawa type interaction.
\begin{equation}
W = \sum_i y_i X \Phi_i \bar{\Phi}_i,
\end{equation} 
It should be noted that in the case of BM model, this $X$ could
be identified with $S$.
Coupling of $X$ to the hidden sector gives rise to vacuum expectation 
values $\vev{X}$ and $\vev{F_X}$ for its scalar and
auxiliary parts respectively. As such the fermionic and scalar parts 
of the messenger sector get masses,
\begin{equation}
m^2_f = |y_i \vev{X}|^2, \qquad 
m^2_s = |y_i \vev{X}|^2 \pm |y_i \vev{F_X}|
\end{equation} 
Thus, the degeneracy between the fermionic and scalar part of the messenger
sector vanishes. 

Gaugino mass arises due to one loop diagrams 
and is given \cite{Dine:1994vc}
\begin{equation}
M_a = \frac{\alpha_a}{4 \pi} \frac{\vev{F_X}}{\vev{X}}\left( 1 + O(x) \right), \quad (a = 1,2,3)   
\label{eq:ginoMass} 
\end{equation}
where $x=\vev{F_X}/(\lambda \vev{X}^2)$. Masses for the scalars of the SUSY 
model for Left-Right symmetric case arises due to two loop corrections
and is given by
%
\begin{equation}
m_\phi^2 = 2  \left( \frac{\vev{F_X}}{\vev{X}} \right)^2
\left[ \left(\frac{\alpha_3}{4\pi}\right)^2 C_3^\phi 
+  \left(\frac{\alpha_2}{4\pi}\right)^2 (C_{2L}^\phi + C_{2R}^\phi)
+  \left(\frac{\alpha_1}{4\pi}\right)^2 C_1^\phi 
\right]\left( 1 + O(x) \right).
\label{eq:SclrMs} 
\end{equation} 
The $C_a^\phi$ are the Casimir group theory invariants defined by,
\begin{equation}
C_a^\phi \delta_i^j = \left( T^a T^a \right)_i^j,
\end{equation} 
where $T^a$ is the group generator of the group which acts on the scalar
$\phi$. Since we consider both the models(ABMRS and BM), the values
of $C_a^\phi$s for the fields are given by,
\begin{eqnarray}
C_3^\phi &=&
\left\{
\begin{array}{ll}
4/3 & \textrm{ for } \phi = Q, Q_c
\nonumber \\
0 & \textrm{ for }\phi = \Phi_i, \Delta's, \Omega's, L's, S,
\end{array} \right.
\nonumber \\ 
C_{2L}^\phi &=&
\left\{ 
\begin{array}{ll}
3/4 & \textrm{ for } \phi = Q, L, \Phi_i \\ 
2 & \textrm{ for } \phi = \Delta, \bar{\Delta}, \Omega\\
0 & \textrm{ for } \phi = Q_c, L_c, \Delta_c, \bar{\Delta}_c, \Omega_c, S,
\end{array} \right.
\nonumber \\
C_{2R}^\phi &=&
\left\{ 
\begin{array}{ll}
3/4 & \textrm{ for } \phi = Q_c, L_c, \Phi_i \\ 
2 & \textrm{ for } \phi = \Delta_c, \bar{\Delta}_c, \Omega_c \\
0 & \textrm{ for } \phi = Q, L, \Delta, \bar{\Delta}, \Omega,S,
\end{array} \right.
\nonumber \\
C_1^\phi &=& 3 Y^2_\phi/5, \textrm{ for each } \phi 
\textrm{ with } U(1) \textrm{ charge } Y_\phi 
\label{eq:CasimirOps} 
\end{eqnarray}
These contributions, will eventually translate into soft
SUSY breaking terms, and in case of BM the desired effective
potential to produce charge preserving vacuum. However, there
is no signal of global parity breakdown and the problem of
domain walls persists.  In the next section we propose a 
modification  of this standard GMSB to explain parity breaking.

\section{Customized GMSB for Left-Right symmetric models} 
\label{sec:customgmsb}
The differences required between the soft terms of the Left and the
Right sector for the DW to disappear at a temperature $T_D$ as given in 
Table \ref{tab:DWalls} are very small. Reasons for appearance of 
this small asymmetry between the Left and the Right fields is hard to
explain since the original theory is parity symmetric. 
However, we now try to explain the origin of this small difference by
focusing on the hidden sector, and relating it to SUSY breaking.

For this purpose we assume that the strong dynamics responsible for SUSY
breaking also breaks parity, which is then transmitted to the visible sector
via the messenger sector and encoded in the soft supersymmetry breaking terms.
We implement this idea by introducing two singlet fields $X$ and $X'$, 
respectively even and odd under parity.
\begin{equation}
X \leftrightarrow X, \qquad X' \leftrightarrow -X'.
\end{equation} 
The messenger sector superpotential then contains terms
\begin{eqnarray}
W &=& \sum_n \left[ \lambda_n X \left( \Phi_{nL} \bar{\Phi}_{nL} 
+ \Phi_{nR} \bar{\Phi}_{nR}\right) \right.
\nonumber \\ &&
+ \left. ~\lambda'_n X' \left(\Phi_{nL} \bar{\Phi}_{nL} 
- \Phi_{nR} \bar{\Phi}_{nR} \right) \right]
\end{eqnarray} 
For simplicity, we consider $n=1$. The fields $\Phi_{L}$, $\bar\Phi_{L}$
and $\Phi_{R}$, $\bar\Phi_{R}$ are complete representations of
a simple gauge group embedding the L-R symmetry group. Further we
require that the fields labelled $L$ get exchanged with fields
labelled $R$ under an inner automorphism which exchanges
$SU(2)_L$ and $SU(2)_R$ charges, e.g. the charge conjugation operation 
in $SO(10)$. As a simple possibility we consider the case when
$\Phi_{L}$, $\bar\Phi_{L}$ (respectively, $\Phi_{R}$, $\bar\Phi_{R}$) 
are neutral  under $SU(2)_R$ ($SU(2)_L$). Generalization to other 
representations is straightforward.

As a result of the dynamical SUSY breaking we expect the fields
$X$ and $X'$ to develop nontrivial vev's and $F$ terms and hence
give rise to mass scales
\begin{equation}
\Lambda_X = \frac{\vev{F_X}}{\vev{X}},
\qquad
\Lambda_{X'} = \frac{\vev{F_{X'}}}{\vev{X'}}.
\end{equation} 
Both of these are related to the dynamical SUSY breaking scale $M_S$,
however their values are different unless additional reasons of symmetry
would force them to be identical. Assuming that they are different
but comparable in magnitude we can show that Left-Right
breaking can be achieved simultaneously with SUSY breaking being
communicated. 

In the proposed model, the messenger fermions receive respective mass 
contributions
\begin{eqnarray} 
m_{f_L} &=& |\lambda\vev{X} + \lambda^{\prime}\vev{X^{\prime}}|\\ \nonumber
m_{f_R} &=& |\lambda\vev{X} - \lambda^{\prime}\vev{X^{\prime}}|
 \end{eqnarray} 
while the messenger scalars develop the masses
\begin{eqnarray}
m_{\phi_L}^2 &=&  |\lambda\vev{X} + \lambda^{\prime}\vev{X^{\prime}}|^2
\pm  |\lambda\vev{F_X} + \lambda^{\prime}\vev{F_{X^{\prime}}}| \\ \nonumber
m_{\phi_R}^2 &=&  |\lambda\vev{X} - \lambda^{\prime}\vev{X^{\prime}}|^2
\pm  |\lambda\vev{F_X} - \lambda^{\prime}\vev{F_{X^{\prime}}}|
\end{eqnarray}
We thus have both SUSY and parity breaking communicated through these
particles.

As a result the mass contributions to the gauginos 
of $SU(2)_L$ and $SU(2)_R$ from  both the $X$  and $X'$ fields with 
their corresponding auxiliary parts take the simple form,
\begin{equation}
M_{a_{L}} = \frac{\alpha_a}{4 \pi} 
\frac{\lambda \langle F_X \rangle + \lambda^{\prime}
\langle F_{X^{\prime}} \rangle}{\lambda \langle X \rangle
+ \lambda^{\prime} \langle X^{\prime} \rangle} \left( 1 + O(x_L) \right)
\end{equation}
where
\begin{equation}
x_L = \frac{\lambda\langle F_X \rangle+\lambda^{\prime}
\langle F_{X'}\rangle}
{|\lambda \langle X \rangle + \lambda^{\prime}\langle X'\rangle|^2}
\end{equation}
and
\begin{equation}
M_{a_{R}} = \frac{\alpha_a}{4 \pi}
\frac{\lambda \langle F_X \rangle - \lambda^{\prime}
\langle F_{X^{\prime}} \rangle}{\lambda \langle X \rangle
- \lambda^{\prime} \langle X^{\prime} \rangle} \left( 1 + O(x_R) \right)
\end{equation}
with
\begin{equation} 
x_R = \frac{\lambda\langle F_X \rangle - \lambda^{\prime}
\langle F_{X'}\rangle}
{|\lambda \langle X \rangle - \lambda^{\prime}\langle X'\rangle|^2}
\end{equation} 
Here $a = 1, 2, 3$. In turn there is a modification to  
scalar masses, through two-loop corrections, expressed to leading
orders in the $x_L$ or $x_R$ respectively, by  the generic formulae
\begin{equation}
m^2_{\phi_{L}} =
2 \left( \frac{\lambda \langle F_X \rangle + \lambda^{\prime}
\langle F_{X^{\prime}} \rangle}{\lambda \langle X \rangle
+ \lambda^{\prime} \langle X^{\prime} \rangle}\right)^2
\left [ \left (\frac{\alpha_3}{4\pi}\right )^2 C_3^\phi +
\left (\frac{\alpha_2}{4 \pi}\right )^2
(C_{2L}^\phi)
 + \left (\frac{\alpha_1}{4 \pi}\right )^2 C_1^\phi \right ]
\end{equation}

\begin{equation}
m^2_{\phi_{R}} =
2 \left( \frac{\lambda \langle F_X \rangle - \lambda^{\prime}
\langle F_{X^{\prime}} \rangle}{\lambda \langle X \rangle
- \lambda^{\prime} \langle X^{\prime} \rangle}\right)^2
\left [ \left (\frac{\alpha_3}{4\pi}\right )^2 C_3^\phi +
\left (\frac{\alpha_2}{4 \pi}\right )^2
(C_{2R}^\phi)
 + \left (\frac{\alpha_1}{4 \pi}\right )^2 C_1^\phi \right ]
\label{eq:ModSclrMs2} 
\end{equation}
where, the values of $C_i^\phi$'s are given by Eqn. \!(\ref{eq:CasimirOps}).  
Applying these formulae to $\Delta$ and $\Omega$, the
parameters $m_i^2$'s and $\beta_i$'s appearing in eqn.s \!(\ref{eq:eqnsoftone}) 
and \!(\ref{eq:eqnsoftthree}) respectively can be calculated. The difference 
between the mass squared of the left and right 
sectors are obtained as
\begin{eqnarray}
\delta m_\Delta^2 &=& 2 \left[ \left( \frac{\lambda \langle F_X \rangle
 + \lambda^{\prime}
\langle F_{X^{\prime}} \rangle}{\lambda \langle X \rangle
+ \lambda^{\prime} \langle X^{\prime} \rangle}\right)^2
- \left( \frac{\lambda \langle F_X \rangle - \lambda^{\prime}
\langle F_{X^{\prime}} \rangle}{\lambda \langle X \rangle
- \lambda^{\prime} \langle X^{\prime} \rangle}\right)^2 \right]
\left\{  \left(\frac{\alpha_2}{4\pi}\right)^2 
+ \frac{6}{5} \left( \frac{\alpha_1}{4 \pi} \right)^2 \right\}
\nonumber\\
&=& 2 (\Lambda_X)^2 \left[ \left( \frac{ 1 + \textrm{tan}\gamma}{1+
\textrm{tan} \sigma}\right)^2
- \left( \frac{ 1 - \textrm{tan} \gamma}{ 1 -
 \textrm{tan} \sigma}\right)^2 \right]
\left\{  \left(\frac{\alpha_2}{4\pi}\right)^2
+ \frac{6}{5} \left( \frac{\alpha_1}{4 \pi} \right)^2 \right\}
\label{eq:Dm2Delta}
\end{eqnarray}
where we have brought $\Lambda_X$ out as the representative mass scale
and parameterised the ratio of mass scales by introducing 
\begin{equation}
\textrm{tan}\gamma = \frac{\lambda^{\prime} 
\langle F_{X^{\prime}} \rangle}{\lambda \langle F_X \rangle},
\quad \textrm{tan} \sigma = \frac{\lambda^{\prime} 
\langle X^{\prime}\rangle}{\lambda \langle X \rangle}
\end{equation}
Similarly, 
\begin{equation}
\delta m_\Omega^2 = 2 (\Lambda_X)^2 \left[
\left( \frac{ 1 + \textrm{tan}\gamma}{1+
\textrm{tan} \sigma}\right)^2
- \left( \frac{ 1 - \textrm{tan} \gamma}{ 1 -
 \textrm{tan} \sigma}\right)^2 \right]
\left( \frac{\alpha_2}{4\pi} \right)^2
\label{eq:Dm2Omega}
\end{equation}
In the models studied here, the ABMRS model will have contribution
from both the above kind of terms. The BM model will have contribution
only from the $\Delta$ fields.

The contribution to lepton masses is also obtained from eq. (\ref{eq:ModSclrMs2}).
This can be used to estimate the magnitude of the overall scale $\Lambda_X$
to be $\geq 30$ TeV \cite{Dubovsky:1999xc} from collider limits. 
Returning to cosmology requirements, we see from table \ref{tab:DWalls} that 
the required mass-squared differences are rather small, and hence it 
is required to assume that ratios $\tan \gamma$ and $\tan \sigma$ must 
be small. Under this assumption,
we expand the quantity in square brackets in eq.s (\ref{eq:Dm2Delta}) and (\ref{eq:Dm2Omega}) 
in small quantities to find its magnitude $4(\gamma -\sigma)$ $+O(\gamma^2, \sigma^2)$. 
The resulting requirement on the values this small parameter is
shown in Table \ref{tab:lambdasp} for various values of $T_D$, the
scale at which the DW finally disappear.

\begin{table}
\begin{center} 
{\setlength\extrarowheight{1.5mm}
\begin{tabular}{cc|c|c|c} 
\hline
$T_D/$GeV & $\sim$ & $10$ & $10^2$ & $10^3$ \\ \hline \hline
Adequate $(m^2-m'^2)$ & $\phantom{\sim}$
& $ 10^{-8} $  & $ 10^{-4} $  & $O(1)$\\ 
Adequate $(\beta_1-\beta_2)$  & $\phantom{\sim}$
& $10^{-12}$ & $ 10^{-8}$ & $10^{-4}$ \\ [1mm] \hline \hline
\end{tabular}}
\end{center}
\caption{Entries in this table are the values of the parameter $\gamma-\sigma$, 
assuming it is small, required to ensure  wall disappearance at temperature $T_D$
displayed in the header row. The table
should be read in conjuction with table \ref{tab:DWalls},
with the rows corresponding to each other. In the last column
first row, $\gamma$ and $\sigma$ are not assumed to be small,  
and the entry signifies the value of the square bracket 
in eq. (\ref{eq:Dm2Delta}), which is generically a quantity $\sim O(1)$. }
\label{tab:lambdasp}
\end{table}

\section{Conclusion}
\label{sec:cnclsn} 
In Left-Right symmetric models, domain walls generically arise in cosmology.
It is necessary to assume the presence of dynamics that eventually signals
departure from exact Left-Right symmetry.
In the absence of such a dynamics the universe would remain
trapped in an unacceptable phase. We have explored models where the scale 
of Left-Right symmetry and the accompanying gauged $B-L$ symmetry are both
low, within a few orders of magnitude of the electroweak scale. In earlier
work we have obtained bounds on the parity breaking parameters so that
domain walls do not conflict with phenomenology, at the same time
providing mechanisms for leptogenesis as well as weak 
inflation\cite{Lyth:1995ka,Matsuda:2000mb}.
The latter is an  effective ways of diluting the density of unwanted 
relics\cite{Kawasaki:2004rx}. 
Wall disappearance is a nonadiabatic phenomenon and could leave behind 
imprints on primordial gravitational wave background
\cite{Grojean:2006bp}. 

The possibilities considered in this paper fall into two categories, whether
weak inflation is permitted or not. Leptogenesis is permitted in all the
cases considered here. Weak inflation becomes possible if the domain
walls linger around for a substantial time, dominating the energy
density of the universe for a limited period. The walls are long lived if
the pressure difference across the walls is small, as happens if
the parity breaking effects are small and difference in effective
potential across the walls is small.

In this paper we have explored the possibility
that smallness of the parity breaking effect is related to the indirect
supersymmetry breaking effects. To be specific we have studied two
viable implementations of Left-Right symmetry, the ABMRS and BM models
discussed in section \ref{sec:mslrm}, and studied them in the context
of gauge mediated supersymmetry breaking as the mechanism. We have
explored a variant of the latter mechanism in order to achieve
parity breaking to be signalled from within the dynamical supersymmetry
breaking sector. Dynamical symmetry breaking effects are associated
with strong dynamics and parity is usually susceptible to breaking
in their presence. 

Our implementation of GMSB contains two ringlets $X$, even under 
parity and $X'$ odd under parity  in the hidden sector  and coupled 
to messengers. Obtaining small parity breaking effect in this context 
generically requires that the scales of the vev's of $X$ and $X'$ be
finely tuned to the same value as seen in table \ref{tab:lambdasp}. 
While this is not unnatural since the strong dynamics is unknown, 
it raises the need to explore the  latter further. In this case all 
the cosmological requirements can be met. 

From table \ref{tab:DWalls}  we see that given that the same dynamics  
determines the soft terms in $\Delta$ and $\Omega$ effective lagrangians, 
the lower bound on $T_D$ required by $\Omega$ fields is higher, thus determining 
the $T_D$. 
On the other hand from table \ref{tab:lambdasp}, we see that given a desirable
value of $T_D$ the terms in $\Delta$ lagrangian require less fine tuning
than the those in the $\Omega$ lagrangian. Since in the BM model the
singlet does not signal any new mass scale, it is
the scale of $\Delta$ vev's which determines the $T_D$. For this reason
it would be more natural model from the point of view of cosmology.

The ABMRS model becomes equally good a contender if for some reason
the $\Omega$ vev scale does not enter the wall disappearance mechanism.
One way this could occur is if the SUSY breaking effects were communicated
primarily to the $\Delta$ sector but not the $\Omega$ sector. In a
class of models considered in \cite{Mohapatra:1997kv, Chacko:1997en} SUSY 
breaking gets communicated by fields charged only under $B-L$ and no other
charges. Since the $\Omega$ are neutral under $B-L$ they would
receive the SUSY breaking effects only as higher order effects.
In this category of SUSY breaking models ABMRS would require less
fine tuning to ensure solution of all the cosmological issues studied here.

\section{Acknowledgement}
We thank K. S. Babu and G. Senjanovic for reading the manuscript and making
useful comments. The work of AS is supported by a CSIR grant.

\appendix
\section{Vacua in the BM model}\label{sec:ap1}
Here we discuss some details of the minimization conditions for the BM model.
We begin with the full renormalizable superpotential for the BM model
\begin{eqnarray} 
W_{LR} &=& {\bf h}_l^{(i)} L^T \tau_2 \Phi_i \tau_2 L_c 
 + {\bf h}_q^{(i)} Q^T \tau_2 \Phi_i \tau_2 Q_c\nonumber\\ 
&& + ~ i {\bf f^\ast} L^T \tau_2 \Delta L 
 + ~i {\bf f} L^{cT}\tau_2 \Delta_c L_c \nonumber \\
&& + ~ S~[~\lambda^\ast ~ {\rm Tr}\,\Delta  \bar{\Delta}
  + ~ \lambda ~ {\rm Tr}\,\Delta_c  \bar{\Delta}_c
  + ~ \lambda^\prime_{ab} {\rm Tr}\, \Phi_a^T \tau_2
 \Phi_b \tau_2 - M_R^2~]\nonumber\\
&& + ~M_\Delta  {\rm  Tr}\, \Delta \bar{\Delta} 
  + M_\Delta^\ast  {\rm Tr}\,\Delta_c \bar{\Delta}_c\nonumber\\
&& + ~\mu_{ab} {\rm Tr}\, \Phi^T_a \tau_2 \Phi_b \tau_2
   + M_s S^2 + \lambda_s S^3
 \end{eqnarray} 
The resulting expressions for the $F$ terms are,
\begin{equation} 
F_s = -[ \lambda^\ast {\rm Tr}\,\Delta  \bar{\Delta}
  + ~ \lambda ~ {\rm Tr}\,\Delta_c  \bar{\Delta}_c
  - M_R^2 + 2M_s S + 3\lambda_s S^2]^\ast
 \end{equation} 
\begin{equation} 
F_\Delta = -[ S\lambda^\ast \bar{\Delta} + M_\Delta \bar{\Delta}]^\ast
 \end{equation} 
\begin{equation} 
F_{\bar{\Delta}} = -[S\lambda^\ast\Delta + M_\Delta{\Delta}]^\ast
 \end{equation} 
\begin{equation} 
F_{\Delta_c} = -[S\lambda\bar{\Delta}_c +  M_\Delta^\ast\bar\Delta_c]^\ast
 \end{equation} 
\begin{equation} 
F_{\bar{\Delta}_c} = -[S\lambda\Delta_c +  M_\Delta^\ast\Delta_c]^\ast
\end{equation} 
From these we assemble the potential for the scalar fields in 
standard notation,
\begin{eqnarray} 
V &=& \left|F_s \right|^2 + \left| F_\Delta \right|^2 +
\left|F_{\bar{\Delta}} \right|^2 + \left|F_{\Delta_c}\right|^2 + 
\left| F_{\bar{\Delta}_c} \right|^2 \nonumber\\
&=& \left| \lambda^\ast {\rm Tr}\,\Delta  \bar{\Delta}
  + ~ \lambda ~ {\rm Tr}\,\Delta_c  \bar{\Delta_c}
 - M_R^2 + 2M_s S + 3\lambda_s S^2 \right|^2\nonumber\\
&& + \left|S\lambda^\ast + M_\Delta \right|^2
 {\rm Tr}\,\bar{\Delta}  \bar{\Delta}^\dagger 
+ \left|S\lambda^\ast + M_\Delta \right|^2
 {\rm Tr}\,\Delta \Delta^\dagger\nonumber\\
&& + \left|S\lambda + M_\Delta^\ast \right|^2
{\rm Tr}\,\bar{\Delta_c}\bar{\Delta_c}^\dagger +
\left|S\lambda + M_\Delta^\ast \right|^2
{\rm Tr}\,\Delta_c {\Delta_c}^\dagger\nonumber\\
&=& \left|\lambda^\ast v_L \bar{v}_L + \lambda v_R \bar{v}_R
 - M_R^2 + 2M_s S + 3 \lambda_s S^2\right|^2\nonumber\\
&& + \left|S\lambda^\ast + M_\Delta \right|^2 
\left(\left|\bar{v}_L\right|^2 + \left|v_L\right|^2\right)
 + \left|S\lambda + M_\Delta^\ast \right|^2 
\left(\left|\bar{v}_R\right|^2 + \left|v_R\right|^2 \right)
\end{eqnarray}
The resulting minimization conditions for the vev's are
\begin{eqnarray}
\frac{\delta V}{\delta S} &=&
\left(2M_s + 6 \lambda_s S\right) Q^\ast + \lambda^\ast
( S\lambda^\ast + M_\Delta)^* \left(\left|\bar{v}_L\right|^2
 + \left|v_L\right|^2\right) \nonumber\\
&& + \lambda\left(S\lambda+ M_\Delta^*\right)^*
\left(\left|\bar{v}_R\right|^2 + \left|v_R\right|^2 \right) = 0
\end{eqnarray}
\begin{equation} 
\frac{\delta V}{\delta v_L} = \lambda^\ast\bar{v}_L Q^*  +
\left|S\lambda^\ast + M_\Delta \right|^2 v_L^* = 0 
 \end{equation} 
\begin{equation} 
\frac{\delta V}{\delta \bar{v}_L} = \lambda^\ast v_L Q^*
+ \left|S\lambda^\ast + M_\Delta \right|^2\bar{v}_L^* = 0
 \end{equation} 
\begin{equation} 
\frac{\delta V}{\delta v_R} = \lambda\bar{v}_R Q^* + 
\left|S\lambda + M_\Delta^* \right|^2 v_R^* = 0
 \end{equation} 
\begin{equation} 
\frac{\delta V}{\delta \bar{v}_R} = \lambda v_R Q^* + 
\left|S\lambda + M_\Delta^* \right|^2 \bar{v}_R^*  = 0
 \end{equation} 
where 
\begin{equation} 
Q = \lambda^\ast v_L \bar{v}_L + \lambda v_R \bar{v}_R
 - M_R^2 + 2M_s S + \lambda_s S^2
\end{equation} 
Thus the desired class of vacua eq. (\ref{eq:SSvev}) is obtained 
provided we ignore the $W^{(2)}$ of eq. (\ref{eq:Wtwo}) in the text and 
choose $\vev{S}$ to be zero.


\end{document}